\documentclass[12pt,pra,aip]{revtex4}
\usepackage[caption = false]{subfig}
\usepackage{graphics,graphicx}
\usepackage{amsfonts}
\usepackage{bm,bbm}
\usepackage{float}
\usepackage{comment}
\usepackage{sidecap}
\usepackage{mathrsfs}
\usepackage{calc}
\usepackage{epsfig}
\usepackage{float}
\usepackage{color}
\usepackage{amsmath}
\usepackage{epstopdf}

\def\bpmat{\begin{pmatrix}}
\def\epmat{\end{pmatrix}}
\def\bmat{\begin{matrix}}
\def\emat{\end{matrix}}

\def\1{\mbox{1\hskip-.25em l}}

\def\beq{\begin{eqnarray}}
\def\eeq{\end{eqnarray}}
\def\be{\begin{equation}}
\def\ee{\end{equation}}

\def\beq{\begin{equation}}
\def\eeq{\end{equation}}

\begin{document}
\title{Atomic and molecular complex resonances from real eigenvalues using standard (hermitian) electronic structure calculations \footnote{Dedicated to the memory of  Howard Taylor who loved science and people}}
\author{Arie Landau$^{1}$, Idan Haritan$^{1,2}$\footnote{The first two authors  contributed equally to this work},  Petra Ruth Kapr\'alov\'a-\v Z\v d\'ansk\'a$^{3}$ and Nimrod Moiseyev$^{1,2,4}\footnote{nimrod@technion.ac.il}$}
\affiliation{$^{1}$Schulich Faculty of Chemistry, $^{2}$RBNI- Russell Berrie Nanotechnology Institute
and $^{4}$Faculty of Physics and the Solid State Institute, Technion - Israel Institute of Technology, Haifa 32000 Israel \\ $^3$
3. Department of Radiation and Chemical Physics at the Institute of Physics, Na Slovance 2, 18221 Prague, and J. Heyrovsky Institute of Physical Chemistry, Dolejskova 3, 18223 Prague, Czech Republic.}
\date{\today}

\clearpage

\begin{abstract}
Complex eigenvalues, resonances, play an important role in large variety of fields in physics and chemistry. For example, in cold molecular collision experiments and electron scattering experiments, autoionizing and pre-dissociative metastable resonances are generated. However, the computation of complex resonance eigenvalues is difficult, since it requires severe modifications of standard electronic structure codes and methods. Here we show how resonance eigenvalues, positions and widths, can be calculated using the standard, widely used, electronic-structure packages. Our method enables the calculations of the complex resonance eigenvalues by using analytical continuation procedures (such as Pad\'{e}).
The key point in our approach is the existence of narrow analytical passages from the real axis to the complex energy plane. In fact, the existence of these analytical passages relies on using finite basis sets. These passages become narrower as the basis set becomes more complete, whereas in the exact limit, these passages to the complex plane are closed.

As illustrative numerical examples we calculated the autoionization resonances of helium, hydrogen anion and hydrogen molecule. We show that our results are in an excellent agreement with the results obtained by other theoretical methods and with available experimental results.
\end{abstract}
\maketitle

\section{Introduction}

\subsection{Motivation}

One of the biggest challenges of electronic structure calculations is to take autoionization into consideration.
Autoionization is a process in which an electronic metastable state decays through spontaneous emission of an electron. Such a state has a finite lifetime, and is known as a resonance. The possibility that in a specific geometrical structure the molecule can be autoionized implies that the electronic and the nuclei coordinates are coupled.
In such a case, the Born-Oppenhiemer (BO) approximation breaks down completely, making the electronic structure computation much harder. A possible solution to the problem is to impose outgoing boundary conditions on the eigenfunctions of the time independent electronic Hamiltonian within the framework of the BO approximation.
In this way, complex potential energy surfaces (CPESs) are obtained. The real and the imaginary parts of the CPES provide, respectively, the energy (position) and the autoionization decay rate (width or inverse lifetime) of a molecule as function of its geometry.
CPESs can be obtained by using complex basis functions \cite{mccurdy_rescigno1978}, analytical continuation of the Hamiltonian's matrix elements \cite{moiseyev_corcoran1979} or one of the complex scaling transformations, such as the uniform \cite{BalsevCombes1971}, exterior \cite{simon1979exterior} or smooth exterior scaling \cite{Nimrod_book}. Alternatively, CPESs can be obtained by introducing a complex absorbing potential (CAP) \cite{Lenz2002rev,muga2004cap} or a reflection-free CAP (RF-CAP) to the Hamiltonian \cite{sajeev2007rf}.

Essentially, the complex electronic eigenvalues obtained within the BO picture serve as potentials of the nuclear time-independent Schr\"{o}dinger equation.
In other words, by using the complex electronic eigenvalues there is no need to go beyond the BO approximation, since they introduce couplings between the nuclear and the electronic coordinates in a simple way.
Still, although this idea was presented many years ago \cite{NM:1981}, so far it did not become a useful approach in the dynamics study of polyatomic molecules with many electrons.
Unfortunately, the reason for this lies in the fact that most of the commercially used codes do not support the above modifications, rendering the computation of complex potential energy surfaces an unconventional task.
However,  the need for reliable CPESs is a must in large variety of fields in chemistry and molecular physics, in which quantum mechanical dynamics of molecules is of interest.
A good example for this acute need is the most recent cold chemistry \cite{edi:2014} and electron scattering experiments \cite{strasser:2014}.
Recently there were serious efforts to develop codes for calculating CPESs \cite{ksenia2013complex,AIK:2013a,AIK:2014complex,white_MHG_mccurdy2015,sajeev2007rf,bala2012br}.
Yet, it is most desirable to have a simple approach that uses standard electronic structure codes in order to calculate CPESs, since these codes are highly optimized and very efficient.

Here, we propose a simple method of calculating CPESs, which utilizes standard electronic structure codes without modifying them.
This approach is based on an analytical continuation of results obtained in the real space into the complex plane.
There are many different approaches to carry out analytical continuation from the real to the complex plane.
Several of which are briefly described in the next section.
In our approach a single real eigenvalue obtained from standard stabilization calculations is analytically continued into the complex plane.
This idea is not new and it faced criticism in the past since the whole eigenvalue plot is not an analytical function of the scaling parameter \cite{mccurdy1983}.
The transition from a stabilization plot into the complex plane and the search for a stationary resonance state goes through a singularity point, known as a branch point (BP) \cite{Friedland1980}.
However, the way we implement this idea avoids these problems by not using the whole stabilizations plot but only an analytical part of it.

\subsection{Background}

As mentioned above, complex energy surfaces are the electronic eigenvalues obtained within the BO picture.
Their real and imaginary parts correspond respectively to the energy position and width.
These eigenvalues represent metastable states with a finite lifetime, resonances.
According to the Balslev and Combes theorem, atomic autoionization resonances becomes square integrable functions by applying uniform complex scaling transformation of the form ${\bf r}\to {\bf r} \eta$, where ${\bf r}$ is the electronic coordinates and $\eta$ is the scaling parameter defined as $\eta=\alpha\exp(i\theta)$ (and $\alpha$ and $\theta$ are real) \cite{BalsevCombes1971,Simon1972,Simon1973}.
In order to implement Balslev and Combes theorem in a molecular system, Simon proposed the use of an exterior scaling transformation that avoids the singularities in the BO molecular Hamiltonian \cite{simon1979exterior}.
In both atomic and molecular transformations the Hamiltonian's spectrum is changing:
while the bound states are unaffected and are characterized by real eigenvalues, the continuum states are rotated into the complex plane by an angle of 2$\theta$, i.e., these complex eigenvalues strongly dependent on $\theta$ \cite{Nimrod_book}.
In addition, the autoionization resonances, characterized by complex eigenvalues, appear in the spectrum.
The first time they are exposed is at a critical value of the rotation angle $\theta$ = $\theta_{BP}$.
At this critical value a branch point is obtained in the spectrum of the complex scaled Hamiltonian (see Fig. 3 in Ref.~\cite{Simon1973,footnote_eta}).
However, as $\theta$ increases the autoionization resonances stabilize and show low dependency on $\theta$, meaning that stationary points in the complex energy plane are obtained.

There are a few early approaches for calculating the resonances position and width from standard hermitian electronic structure calculations.
For example, resonances can be approximated from the density of states in the continuum, they can be calculated from the asymptotes of the continuum eigenfunctions or from the phase shifts of the eigenfunctions (see chapter 3 in Ref.\cite{Nimrod_book}).
Alternatively, the resonances can be obtained from stabilization calculations where the eigenvalues are computed for an increased number of basis functions 
 \cite{Midtdal1966} or when a finite given basis functions are scaled by a real factor 
 \cite{taylor1971,taylor1976hazi}.

Here propose to analytically dilate the real energies into the complex plane via the Pad\'{e} approximant.
At its basis this approach is also not new: In 1981 Simons used the stabilization calculations and suggested to carry out a unitary transformation from the adiabatic energy levels to the diabatic presentation \cite{simons1981}.
In the diabatic representation the electronic energy levels are coupled, unlike in the adiabatic picture, where the electronic energies are represented as non-interacting states.
Therefore, in the adiabatic picture avoided crossings are obtained, while in the diabatic picture crossings are exposed in the complex plane. These crossings represent the branch points, and their exposer facilitates locating the nearby stationary points (the resonance energy).

An even simpler approach was introduced by Thompson and Truhlar \cite{Truhlar1982} and later modified by Isaacson and Truhlar \cite{Truhlar1984}.
Under this approach the resonance complex energy is obtained by analytically continuing a single eigenvalue into the complex plane, to this end the whole stabilization curve is used.
However, as McCurdy and McNutt \cite{mccurdy1983} pointed out, analytic continuation of a single-root fails due to the existence of non-analytic regions in the eigenvalue plot.
Hence, McCurdy and McNutt suggested a multi-eigenvalue method. They carried out analytical continuation of the characteristic polynomials of the Hamiltonian matrix using at least two eigenvalues.
Doing so, they avoided the non-analyticity of the corresponding eigenvalues and located stationary points.
In fact, McCurdy with McNutt opened a new research direction for calculating resonances from stabilization graphs. This multi-eigenvalue method is indeed pursued until today \cite{jordan1990,jordan2014} and it relies on the correct description of the avoided crossings.

Nevertheless, in this paper we show that under easy to fulfill conditions, resonances can be calculated from a single stabilization root, similar to Thompson and Truhlar \cite{Truhlar1982}.
Yet, unlike Thompson and Truhlar, we suggest to analytically dilate the eigenvalue as function of the stabilization parameter while avoiding the branch point areas, i.e., the non-analytic structure of the complex energy.
This can only be achieved when carefully choosing the area in the stabilization plot that will be dilated, meaning, excluding the avoided crossings. In other words, we are only using the stable analytical region in the stabilization graph.
We are not interested in describing branch points.
Furthermore, we illustrate later that we can always remain in an analytic area, and eventually converge to a stationary point in the complex plane. The converged energy is in an excellent agreement with the eigenvalue obtained by an explicit complex scaling calculations.

The proposed analytical continuation scheme resembles Moiseyev's and Corcoran's suggestion to treat molecular resonances within the BO picture by evaluating the molecular Hamiltonian matrix elements \cite{moiseyev_corcoran1979}.
These matrix elements are analytical functions of the scaling factor, even though the operator is not, because the contour of their integration in the complex plane can be chosen to be such that avoids the singular points \cite{mccurdy_rescigno1978,moiseyev_corcoran1979,mccurdy1980_BO,mccurdy1980_BF,simon_morgan1981}.
Therefore, they can easily be analytically continued to the complex plane \cite{moiseyev_corcoran1979,mccurdy_rescigno1978}.
In a similar way, we demonstrate here the existence of an analytical path from the stabilization graph towards a complex stationary point.
This path bypasses any branch point, thus avoiding any singularity.

\section{Theoretical Scheme}

\subsection{Resonances from square integrable basis functions: complete vs. finite bases}


Upon complex scaling, autoionization resonances become square integrable functions, therefore are exposed as eigenfunctions in the Hilbert space.
When the complex scaling parameter is defined as $\eta=\alpha\exp(i\theta)$, the continuum states are strongly dependent on $\theta$, however, in the exact limit, they are invariant to $\alpha$ \cite{Nimrod_book}.
That is, when using a complete basis set, the continuum states stay fixed as $\alpha$ is varied \cite{BalsevCombes1971,Simon1972,Simon1973,footnote_eta}.

\begin{figure}
\subfloat[]{\includegraphics[width = 3.4in]{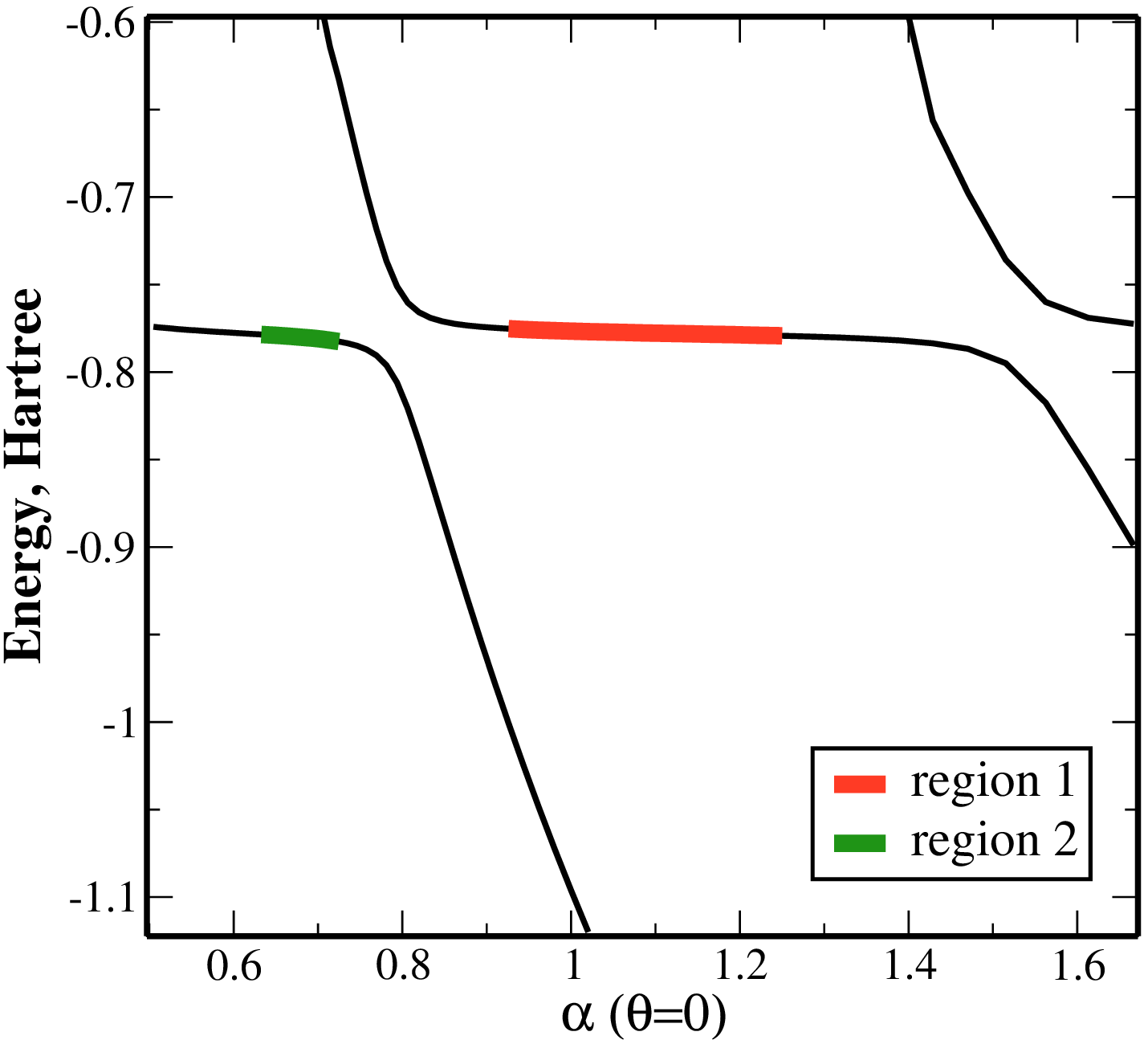}}
\subfloat[]{\includegraphics[width = 3.5in]{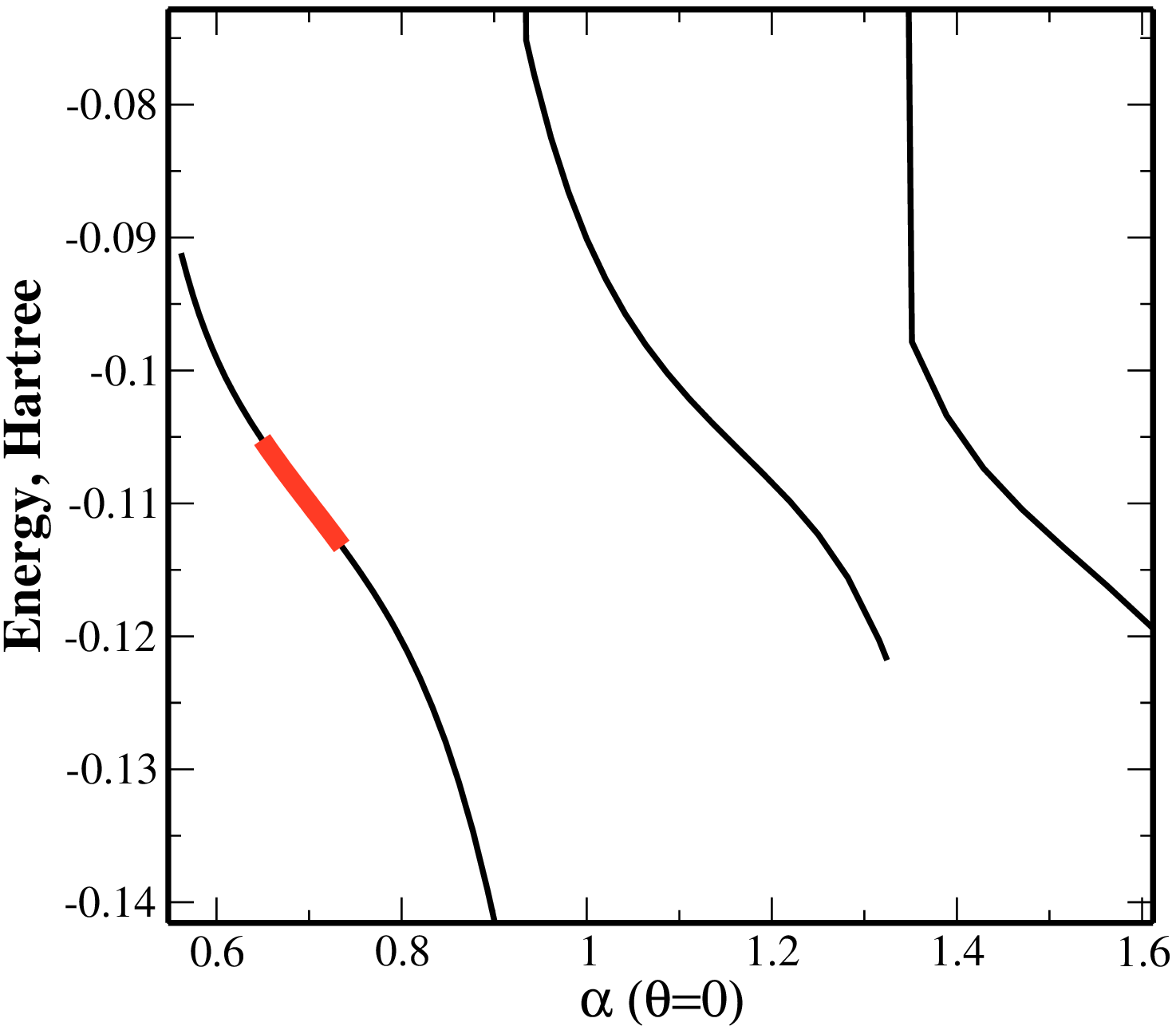}}
\caption{
[Color online] Energy stabilization plots for (a) the helium $2s^2$ resonance state (b) the hydrogen molecular $1\sigma_u^2$ resonance state ($R=1.4$ {\it a.u.}). The stabilization is obtained by varying the real scaling parameter, $\alpha$. The red and green areas in (a) represent the two regions that were used as a starting point for the two analytical continuations. For H$_2$ only one stable region, marked in red in (b), was used for the analytical continuation.}
\label{stabilization}
\end{figure}

Contrary, in actual calculations, when incomplete and finite basis sets are employed, the continuum spectrum does vary with respect to $\alpha$.
An example for this behavior is seen in Fig.~\ref{stabilization}. In this figure the relevant energy levels of the helium atom and the hydrogen molecule are plotted as a function of $\alpha$.
In both cases a finite number of Gaussians were used as a basis set in the calculations (further computational details are given below) and it is clear that the states depend on $\alpha$.
This dependency can be utilized in order to calculate the resonance position and width.
To do so one must first identify the resonance footprints in the real energy spectrum. These footprints lie in the stabilization plot. In Fig.~\ref{stabilization} these footprints are clear: while continuum energies change strongly with respect to $\alpha$, one specific energy seems to stay fixed and emerges out of the plot.
This energy corresponds to the resonance position which is not highly affected by the variation of the scaling factor. The reason for this stabilization is clear: the resonance wave-function is localized in space. Therefore, relatively small expansion or compression of the basis functions does not have much effect on it.
In contrast, the other states are associated with delocalized functions and therefore are strongly affected by the scaling factor.

Although the resonance position emerges out of the stabilization plot, its width does not.
One way to obtain its width is to move into the complex plane.
There, the resonance is square integrable, and appears as a complex eigenvalue of the Hamiltonian. The imaginary part of this eigenvalue corresponds to the resonance width.
One easy method to move into the complex plane is by analytical continuation from the stabilization plot \cite{simons1981,mccurdy1983,Truhlar1982,Truhlar1984,jordan1990,jordan2014}. 
However, this continuation should be done with caution since an avoided crossing in the stabilization plot is associated with singularity, a branch point in the complex plane.
This singularity is caused due to the fact that as the resonance becomes square integrable, it separates from the continuum.
At this critical point, the resonance and the rotating continuum coalesce in the complex plane. This causes the spectrum to be deficient and thus non-analytic.

The existence of an avoided crossing in the stabilization graph raises difficulties, and  analytically continuing a single eigenvalue can fail to accurately find a stationary point in the complex plane \cite{mccurdy1983}.
For this reason, McCurdy and McNutt \cite{mccurdy1983} and later Jordan and coworkers \cite{jordan1990,jordan2014} used at least two eigenvalues to obtain a truncated characteristic polynomial of the Hamiltonian.
Then, they performed analytical continuation of the polynomial coefficients instead of the eigenvalue.
Yet, since finite basis sets are always employed, there is a way to analytical continue a single eigenvalue into the complex plane.
The ``trick'' is to use only a small part of the stabilization plot, a relatively stable region, instead of the whole plot. This region is an analytical function of $\alpha$ and it avoids the avoided crossings, therefore, it is a perfect starting point for the desired analytical continuations.

It is important to stress that this ``trick'' is only applicable in a finite basis set. Under an infinite basis set it is impossible to carry out analytical continuation of an eigenvalue into the complex plane.
As the basis set approaches completeness the number of avoided crossings increases, and thus the singularity area in the complex plane increases too.
That is, there is no way to analytically dilate a single eigenvalue and avoid the branch point.

In the next section, a comprehensive discussion and illustration are presented explaining why when finite basis sets are used, it is possible to bypass the singularity and find an analytical route towards the desired stationary point.
Moreover, it is shown that when finite basis sets are used, it is actually very difficult to find a branch point let alone pass through it.

\clearpage

\begin{figure}
\includegraphics[scale=1.0]{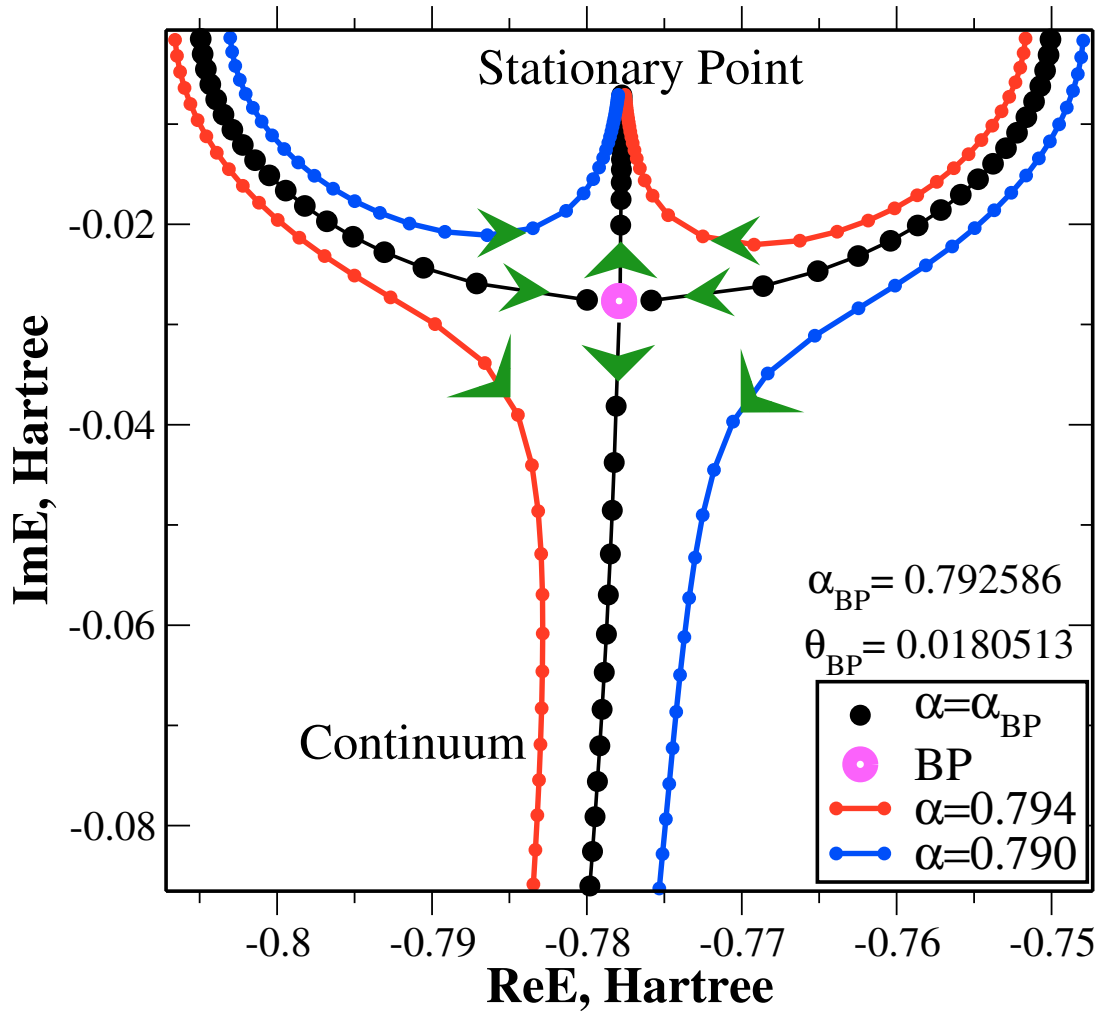}
\caption{[Color online] Complex eigenvalues of a non-hermitian helium Hamiltonian matrix obtained by the use of uniform complex scaling, $\eta=\alpha\exp(i \theta)$ where $\theta$ is varied and $\alpha$ is held fixed.
Three $\theta$-trajectories are presented with different $\alpha$ values.
The $\alpha_{BP}$ trajectory is presented in black and includes also the branch point (pink).
The green arrows represent the direction of the $\theta$-trajectories from zero by 0.001 radians increments.
Note, that the rotational angle $\theta_{BP}$, in which the branch point is obtained, is much smaller than the value of $\theta$ at the stationary point.
In these uniform complex scaling calculations a finite Gaussians basis set is used.}
\label{bp}
\end{figure}

\clearpage

\subsection{A finite basis set as a means to bypass the branch point}

As previously discussed, the resonance position and width can be obtained by moving into the complex plane using the complex scaling parameter $\eta=\alpha\exp(i\theta)$ .
In the complex plane, the resonances are recognized as complex energies that are not affected by a small change in
$\alpha$ and $\theta$, i.e. stationary points.
Graphically, we identify them as cusps in the $\theta$ or $\alpha$-trajectories plots.
Such stationary points are located close to the branch point, but are always deeper into the complex plane (see Fig. 2 in Ref.~\cite{mccurdy1983} and Fig. 2 in Ref.~\cite{Friedland1980}).
Another good example for this is given in Fig.~\ref{bp}, where $\theta$-trajectories calculations for the $2s^2$ resonance state of helium are presented. In these trajectories, $\alpha$ is held fixed and the rotational angle, $\theta$, is increased from zero (the hermitian case).
As indicated in the black $\theta$-trajectory, the rotational angle $\theta$ = $\theta_{BP}$, in which the branch point is obtained, is much smaller than the value of $\theta$ at the stationary point.
That is, in order to locate a stationary point by varying $\theta$ from the real axis, it is necessary to go through a branch point.
Theoretically, this feature is a problem when one tries to find the stationary point through analytical continuation, however as clearly seen in Fig.~\ref{bp}, in practice it is not.
In this figure, two additional trajectories are plotted. These blue and red trajectories reach the stationary point smoothly without passing through a branch point.
In other words, contrary to the complete basis set case, in which the branch point is $\alpha$-independent, when a finite basis set is used, the branch point is obtained for a very specific value of $\alpha$ = $\alpha_{BP}$ (as shown in Fig.~\ref{bp}).
As a matter of fact, it is very hard to calculate a branch point and there are specific methods developed in order to locate it \cite{Friedland1980, raam2010_3points}.
Here for example, the branch point parameters, $\theta_{BP}$ and $\alpha_{BP}$, were found using such a method \cite{raam2010_3points}.
However, these parameters are never fully exact but rather a mere estimate to the real values.
To sum it up, for finite basis sets, the fact that the branch point is obtained only for very specific values of $\alpha$ and $\theta$ is an advantage.

In light of the above, the resonance stationary point can be found via analytical continuation of the real eigenvalue from the stabilization plot into the complex plane.
This is examined below by comparing atomic and molecular resonance positions and widths obtained by the proposed analytical continuation with the results obtained by explicit uniform complex scaling methods.

Below a mathematical demonstration is given for the claim that in a finite basis set analytical continuation to the complex plane is feasible.
It is based on the fact that $\alpha$ is always different than the exact $\alpha_{BP}$.

Sufficiently close to the branch point where the stationary point lies ($\alpha$ = $\alpha_{SP}$), the complex eigenvalues associated with the resonance and continuum states are given by the leading terms in the Puiseux series \cite{Nimrod_book, Friedland1980} -
\begin{eqnarray}
 E_{\pm}&\cong& E_{BP}\pm b\sqrt{{\eta}-{\eta_{BP}}}\nonumber \\
&=&  E_{BP}\pm b\sqrt{\alpha_{SP} e^{i\theta}-\alpha_{BP}e^{i\theta_{BP}}}\nonumber \\
&=& E_{BP}\pm b\sqrt{\eta_{BP}}\bigg(\frac{\alpha_{SP}}{\alpha_{BP}}e^{i\theta-i\theta_{BP}}-1\bigg)^{1/2},
\end{eqnarray}
Here, coalescence of two solutions can only occur when -
\begin{equation}
\Delta(\theta)=\frac{\alpha_{SP}}{\alpha_{BP}}e^{i\theta-i\theta_{BP}}-1\to 0,
\end{equation}
Therefore, it is clear that a coalescence is an unlikely scenario, particularly when $\theta$ = $\theta_{BP}$.
That is, since $\alpha_{SP} \neq \alpha_{BP}$ the eigenvalue associated with the stationary point is an analytical function of $\theta$.

\clearpage

\subsection{Analytical continuation via the Pad\'{e} approximant\\ $-$ from real eigenvalues to the complex energy plane }
\protect\label{sec:pade}

In this paper, we analytically dilate a hermitian eigenvalue to the complex plane by the Pad\'{e} approximant.
More specifically, we generate an analytical approximation to E($\eta$), where $\eta=\alpha\exp(i\theta)$, by the Schlessinger point method \cite{Schlessinger1966}.
This method requires a set of $M$ input variables and their corresponding values.
In our case, we select $\eta$ as the input variable, or more precisely $\eta$ = $\alpha$ (where $\theta$  = 0),
and $E(\alpha)$ as the corresponding value.
The input data, $E(\alpha)$, is taken from the relatively stable region in the stabilization plot which excludes the avoided crossing.
For example, in order to calculate the 2$s^2$ resonance state of helium, we used the red or green regions in the stabilization plot shown in Fig.~\ref{stabilization}(a).
That is, each region (red or green) is used as a starting point for a different calculation.
The Schlessinger truncated continued fraction has the form  -
\begin{eqnarray}
\label{PADE}
C_M(\alpha) = \frac{E(\alpha_1)}{1+\frac{z_1(\alpha-\alpha_1)}{1+\frac{z_2(\alpha-\alpha_2)}{\vdots\, z_M(\alpha-\alpha_M)}}},
\end{eqnarray}
where the $z_i$ coefficients are chosen such that
\begin{equation}
C_M(\alpha_i) =E(\alpha_i),\,\, i= 1,2 ,..., M.
\end{equation}
Once the $z_i$ coefficients are determined, we perform an analytical continuation into the complex plane by evaluating $C_M(\eta)$ (where $\eta$ is complex, i.e. $\theta \neq $ 0). Convergence of the extrapolated function, $C_M(\eta)$, with respect to $M$ is routinely checked, and the difference between $C_M(\eta)$ and $C_{M-1}(\eta)$ is reported as the Pad\'{e} error.

The procedure we propose is summarized as follows:\\
1) Calculate a stabilization plot using a standard electronic-structure approach as in Fig.~\ref{stabilization}.\\
2) Use the points in a stabilized region as input for the Schlessinger point method.
Remember to exclude the avoided crossings like the colored regions in Fig.~\ref{stabilization}.\\
3) Fit the selected points to a Pad\'{e} polynomial by making sure that this polynomial reproduce the original values for the input variables.\\
4) Preform analytical continuation in the real plane (i.e., with $\theta$ = 0) to make sure an appropriate set of points was chosen.
A too dense set of input points will damage the performance of the Pad\'{e} approximant, and will end up in a linear extrapolated plot that is almost independent of $\alpha$.\\
5) Preform analytical continuation into the complex plane:
generate a $\theta$-trajectory with a fixed $\alpha$. The fixed $\alpha$ should be taken from its values in the stable region of the stabilization plot.\\
6) Look for an optimal $\theta$ value in which a cusp is seen in the $\theta$-trajectory. \\
7) Calculate an $\alpha$-trajectory using the optimal $\theta$ found in step 6. Look for the optimal $\alpha$ in which a cusp is seen in the $\alpha$-trajectory. \\
8) Repeat steps 5-7 until the optimal $\alpha$ and $\theta$ are converged.
Upon convergence a clear cusp is obtained in both $\alpha$ and $\theta$-trajectories.
Both cusps will touch each other (see Fig.~\ref{H2-pade}).
These cusps are associated with a stationary point that satisfy
${\partial \over {\partial \alpha}}E(\eta) \rightarrow 0$ and ${\partial \over {\partial \theta}}E(\eta)  \rightarrow 0$ \cite{Nimrod_book}. \\
9) Check the Pad\'{e} errors of the stationary point and make sure it is reasonable.\\
10) It is recommended to repeat the process with a slightly different set of input points to examine the stability of the stationary point. \\

This procedure is very easy and quick. The only  time and computational consuming step is step 1. That is, this procedure is much simpler than actual electronic-structure calculations inside the complex plane.

\section{Computational details}
\protect\label{sec:details}

In this work we study resonance states of the helium atom, the H$^-$ anion and the H$_2$ molecule with the leading 2$s^2$, 2$s^2$ and $1\sigma_u^2$ configurations, respectively.
In all cases we reach the full configuration interaction (FCI) limit.
For helium we use our own CI code for two-electron atoms and ions including complex scaling transformation \cite{petra2013helium}.
For H$^-$ and H$_2$ we use the standard equation-of-motion couple-cluster with singles and doubles (EOM- CCSD) for excitation energies implemented in the {\it Q-Chem} package \cite{qchem_4}.
The ground states, which play the role of the reference states, are 1$s^2$ for helium and H$^-$, and $1\sigma_g^2$ for H$_2$.

The basis sets employed comprise of primitive Gaussian functions since they can be scaled trivially. Uniform complex scaling is performed by rotating the Gaussians basis functions by ${\bf r}\to {\bf r} \eta$, where ${\bf r}$ is the Gaussian coordinate and $\eta$ is the scaling parameter defined as $\eta=\alpha\exp(i\theta)$ ($\alpha$ and $\theta$ are real) \cite{BalsevCombes1971,Simon1972,Simon1973}.
Stabilizations are obtained by scaling the Gaussians when $\theta$ is set to zero.

For helium we use a 19$s$ 15$p$ 10$d$ 8$f$ (19/15/10/8) series of exponentially tempered basis functions \cite{petra2013helium}.
The basis set is optimized for calculating the helium states with the highest principal number $n$=2 and the highest partial angular momentum $l$=3 (1$s^2$, 1$s2s$,  ..., 2$f^2$) for about 10$^{-5}$ Hartree precision.


For H$^-$ we use a 12$s$ 10$p$ 5$d$ 3$f$ 2$g$ (12/10/5/3/2) set based on the Bentley and Chipman basis, which include $s$- and $p$-type functions, and was used to calculate the H$^-$ resonance before \cite{bentley_chipman1987}.
The contracted 1$s$ orbital is transformed into three primitive functions. In addition, the basis is augmented with the $d$-, $f$- and $g$-type functions of the aug-pcJ-4 basis set \cite{jensen2010aug_pcJ}.
Convergence of the resonance energy with respect to the  number of diffuse basis functions is ensure by further augmenting this basis with two $s$- and two $p$-type  diffuse functions, which results in a 14/12/5/3/2 basis set.

For H$_2$ we report results obtained by using a 12$s$ 10$p$ 3$d$ 2$f$ (12/10/3/2)  basis, which is based on the 5ZP basis set \cite{5ZP_2006}.
In this basis we replace the most diffuse $s$-function with six diffuse functions, and the most diffuse $p$-function with seven diffuse functions,  where an even-tempered spacing of 2.0 was employed for their construction. In addition, the most tight $d$- and $f$-type functions where replaced with diffuse functions.
This basis is referred to as Basis1.  Basis set convergence is ensured by comparing this basis set results with two other bases.  First, to make sure we obtain convergence with respect to the number of diffuse basis functions, we compare the 12/10/3/2 results with the ones obtained with a similar basis in which the most diffuse $s$ and $p$ functions are removed (a 11/9/3/2 basis set). We observe convergence within
$10^{-8}$ Hartree for the resonance energy between the two bases.
Second, we examined a denser 12/9/3/2  basis set, in which the three most diffuse $s$-functions of the original 5ZP basis set are replaced with eight diffuse functions (with even-tempered spacing of 2.0), everything else is similar to the 11/9/3/2  basis set. This basis is referred to as Basis2. Below, we report results obtained by Basis1 and Basis2.

All the bases exponents  are presented in the supplementary information.

%


\section{Numerical Applications}

\subsection{Helium autoionization resonance $-$ a comprehensive test case}

\begin{figure}
\subfloat[]{\includegraphics[width = 3.1in]{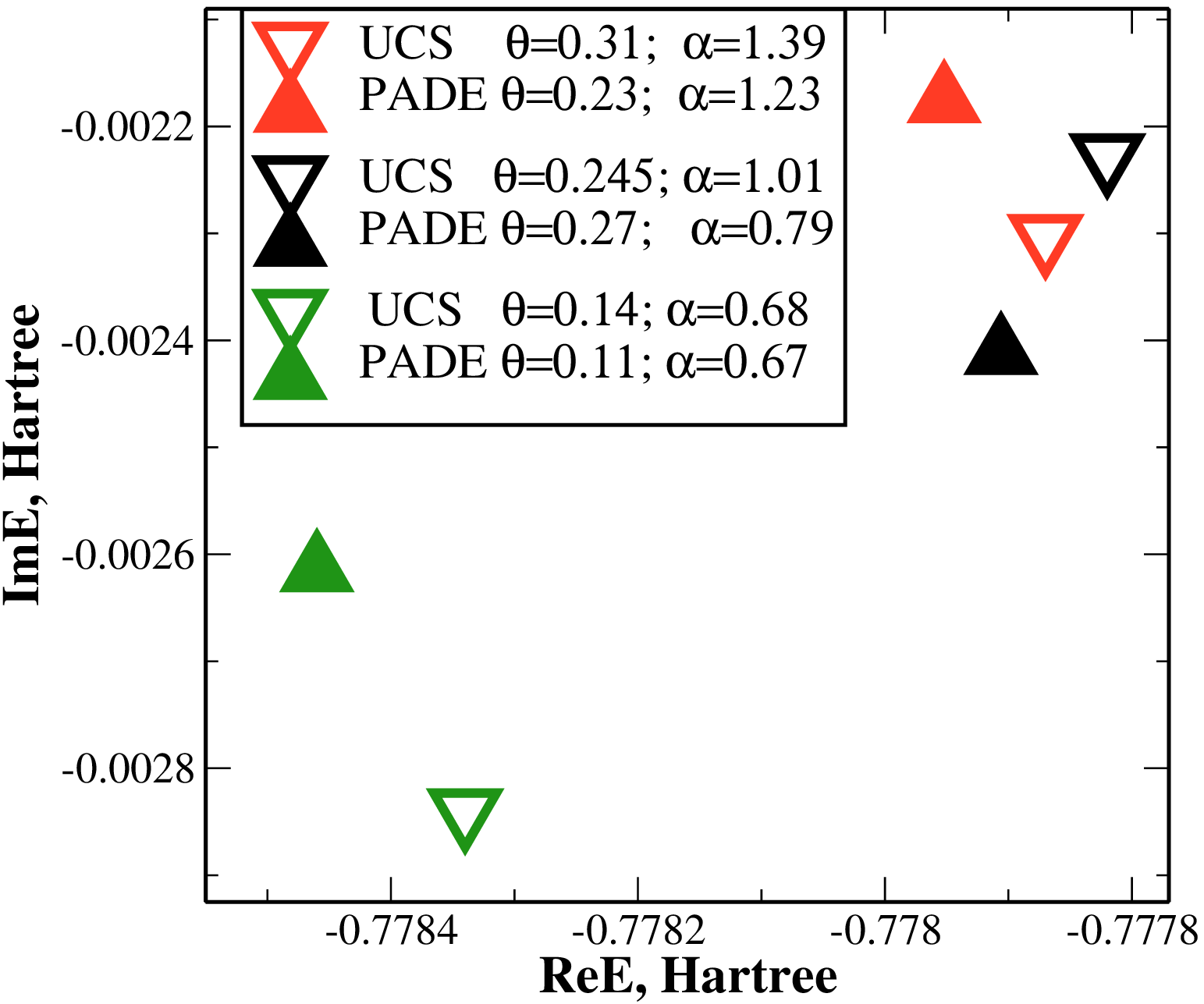}}
\subfloat[]{\includegraphics[width = 3.05in]{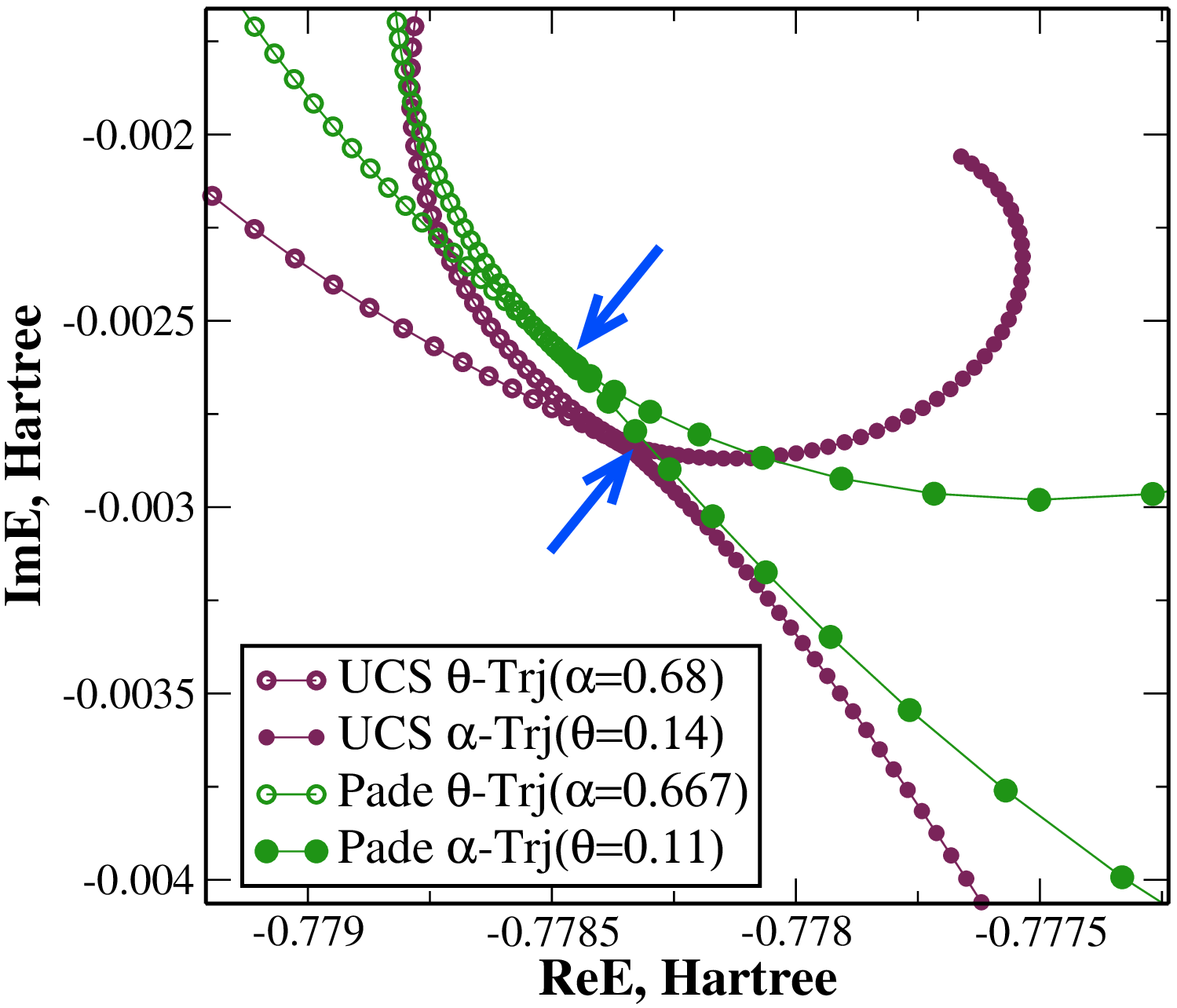}}\\
\subfloat[]{\includegraphics[width = 3.2in]{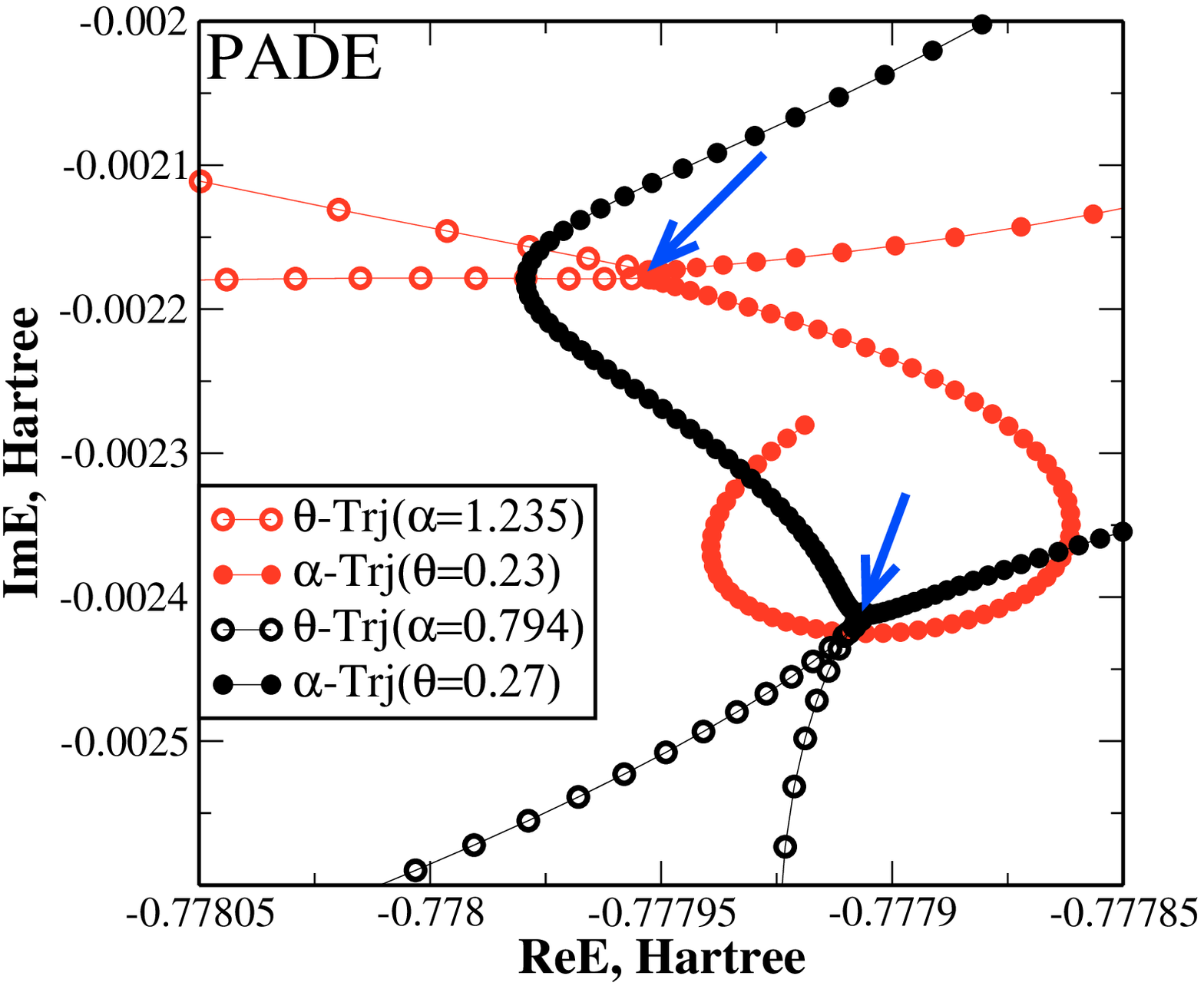}}
\subfloat[]{\includegraphics[width = 3.1in]{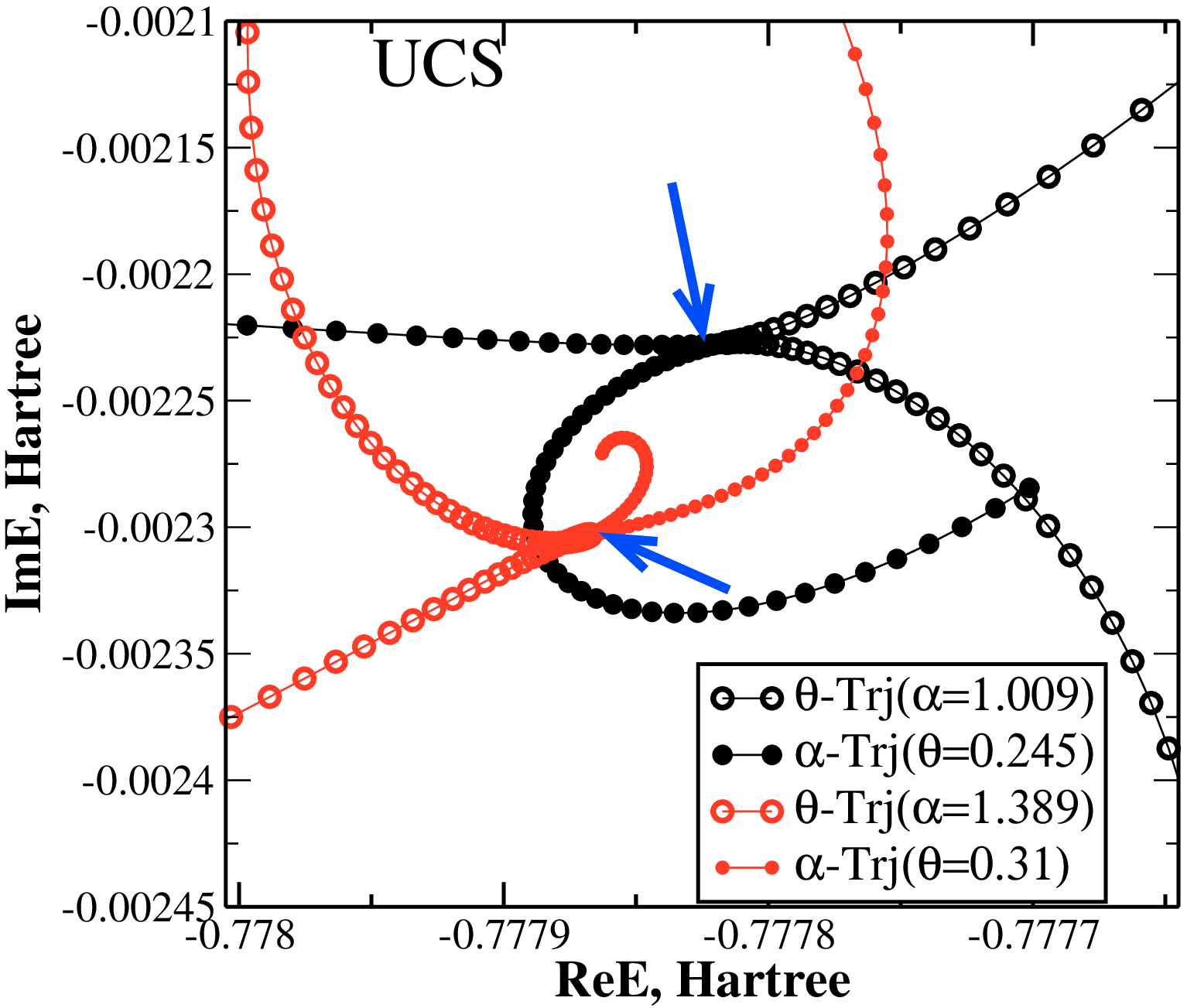}}
\caption{[Color online] Helium $2s^2$ resonance stationary points and the corresponding $\alpha$- and $\theta$-trajectories.
(a) Stationary points energies for the $2s^2$ resonance. Empty triangles represent stationary points obtained by UCS, while the full triangles are stationary points obtained by our analytical continuation. Each triangle correspond to the cusps in (b), (c) and (d).
(b) $\alpha$- and $\theta$-trajectories obtained from our analytical continuation scheme starting from region 2 in Fig.~\ref{stabilization}(a) (green lines) and the corresponding UCS calculations (purple lines).
(c) $\alpha$- and $\theta$-trajectories obtained by our analytical continuation scheme starting from region 1 in Fig.~\ref{stabilization}(a).
(d) $\alpha$- and $\theta$-trajectories obtained from the UCS calculations that correspond to (c).
In (b) a clear cusp is seen, indicating one stationary point.
In (c) and (d) two cusp are seen (one in black and the other in red) indicating two stationary points. Every stationary point is marked with a blue arrow.
In (b) (c) and (d) the closed and open circles correspond to the $\alpha$- and $\theta$-trajectories, respectively.}
\label{cusps}
\end{figure}

Since helium is a well-studied and simple system,~\cite{petra2013helium} the helium $2s^2$ state can serve as a comprehensive test case, for which we can study autoionization processes and explore the capabilities of new computational schemes.
Furthermore, helium is a two electron system, hence it is easy to calculate this resonance position and width using FCI and uniform complex scaling (UCS). 
Doing so, no approximation is done on the electronic structure and a pure comparison to a new method can be performed. Therefore, in our study this calculation was used as a reference point.
Such a comparison for helium $2s^2$ resonance can be seen in Fig.~\ref{cusps}.
In this figure, optimal cusps from both UCS calculations and our analytical continuation calculation are presented.
For the analytical continuation calculations the cusps were obtained through the iterative procedure described in Section~\ref{sec:pade}.
A similar iterative scheme, in which FCI calculations were performed in each step, was carried out for the UCS calculation.
A remarkable agreement between the UCS calculations and our analytical continuation is clearly observed.

In Fig.~\ref{cusps}(b) two cusps are displayed. One cusp was obtained by using region 2 in Fig~\ref{stabilization}(a)  as the starting point for the Pad\'{e} analytical continuation. The converged scaling parameters of the Pad\'{e} cusp were then used as starting points for the UCS iterative calculations. Doing so, a UCS cusp was also obtained. The two cusps were in an excellent agreement, both in their scaling parameters and in their position and width. In other words, the Pad\'{e} approximant was able to produce a very accurate $\alpha$- and $\theta$-trajectories based on the real stabilization plot.

In Fig.~\ref{cusps}(c) two cusps are displayed for the Pad\'{e} analytical continuation. This time, the cusps were obtained using region 1 in Fig~\ref{stabilization}(a)  as the starting point for the Pad\'{e} approximant. Each cusp represented another resonance stationary energy, where the two energies differ by $2\times10^{-4}$ Hartree for the imaginary part, and by $4\times10^{-5}$ Hartree for the real part. Similarly, in Fig~\ref{cusps}(d) two cusps are also displayed for the UCS calculations. Again, each cusp represented another resonance stationary energy where these two energies are within less than $10^{-4}$ Hartree difference for both real and imaginary parts. It is important to note that the UCS cusps were obtained as described above. That is, the converged scaling parameters of the Pad\'{e} cusps were used as starting points for the UCS iterative calculations.

A good summary of Fig.~\ref{cusps}(b-d) is displayed in Fig.~\ref{cusps}(a). This figure demonstrates the proximity of the Pad\'{e} results to the UCS ones. Each stationary point obtained by the UCS calculations has an analogues point obtained by the Pad\'{e} approximant, where in all cases there is a good agreement between them. In fact, the distance between each cusp couple was calculated and was found to be:
$1.4\times10^{-4}$, $1.1\times10^{-4}$ and $2.5\times10^{-4}$ Hartree for the black, red and green cusps, respectively.
Recalling the fact that the Pad\'{e} procedure is a fast and simple computational scheme, these results are very encouraging.

%
%
%

\begin{figure}
\subfloat[]{\includegraphics[width = 0.45\textwidth]{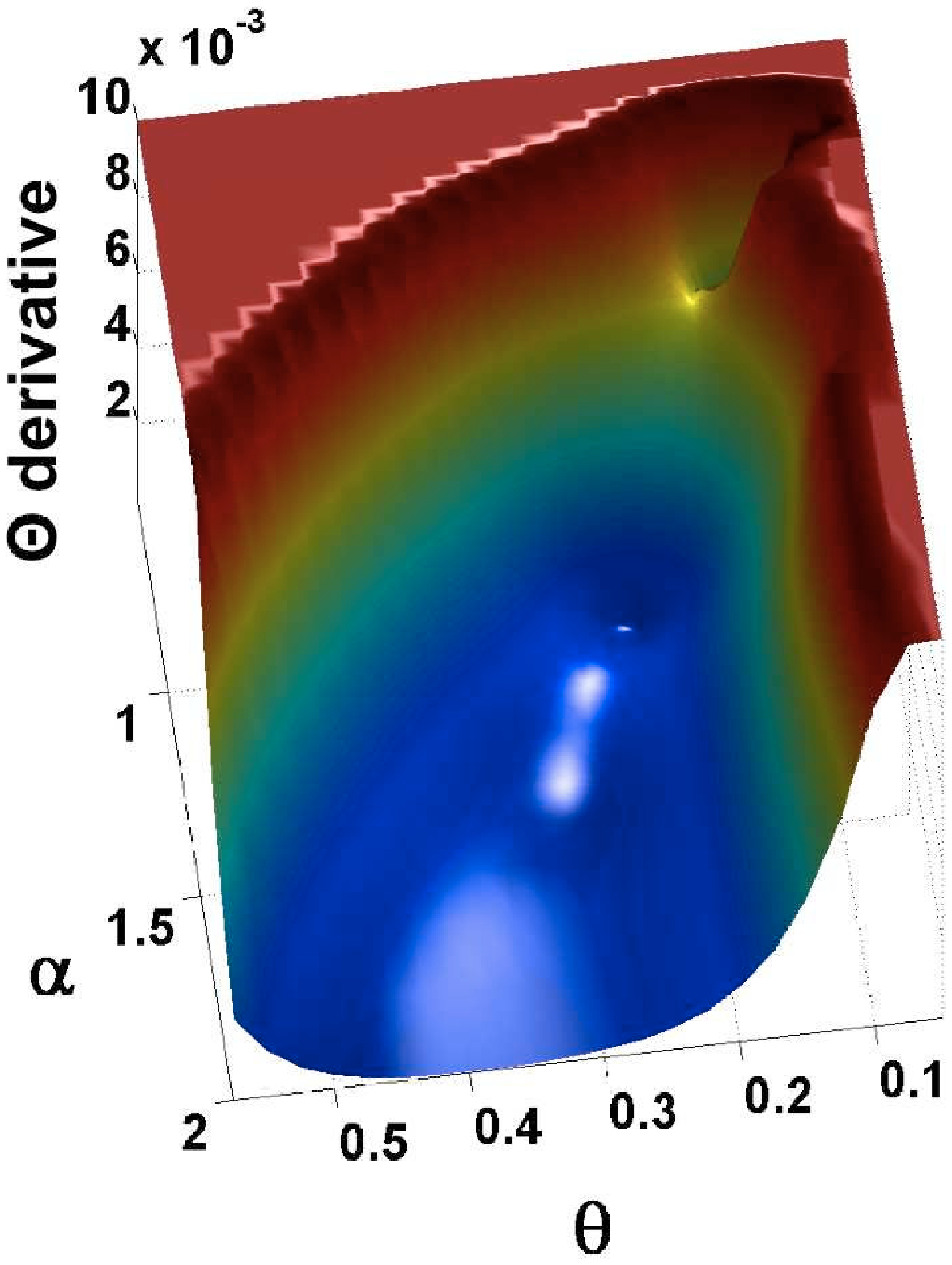}}
\subfloat[]{\includegraphics[width = 0.45\textwidth]{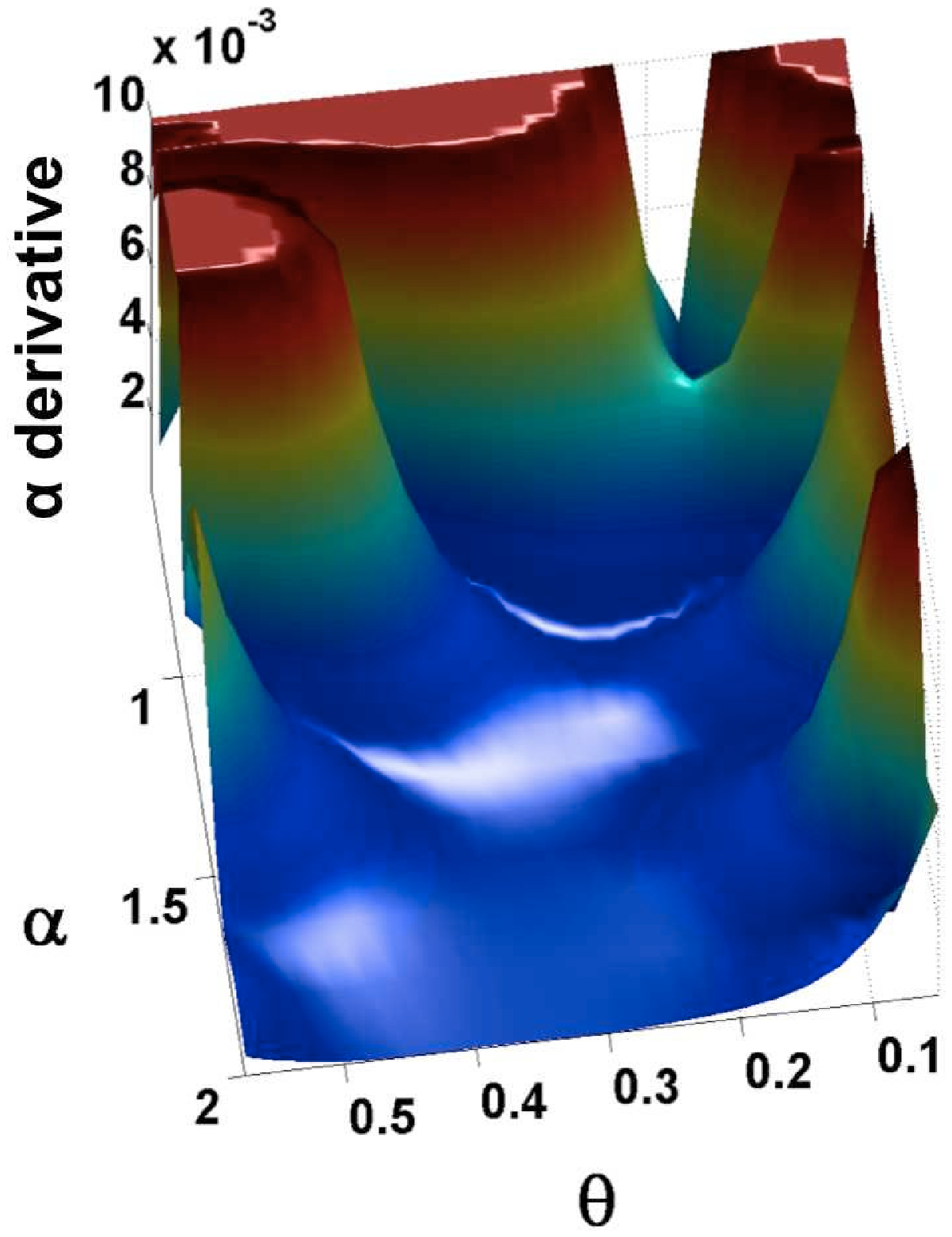}}\\
\subfloat[]{\includegraphics[width = 0.45\textwidth]{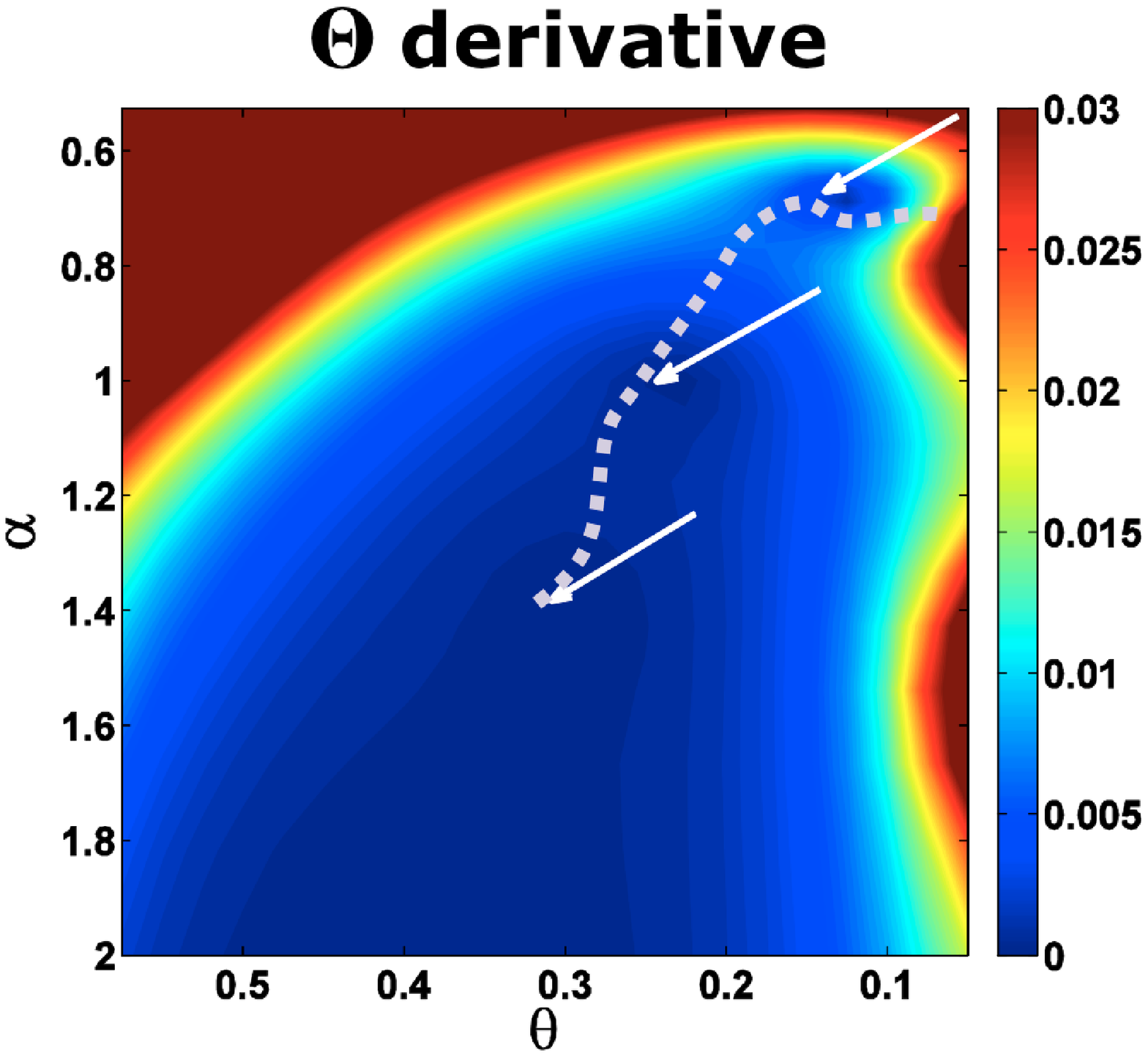}}
\subfloat[]{\includegraphics[width = 0.45\textwidth]{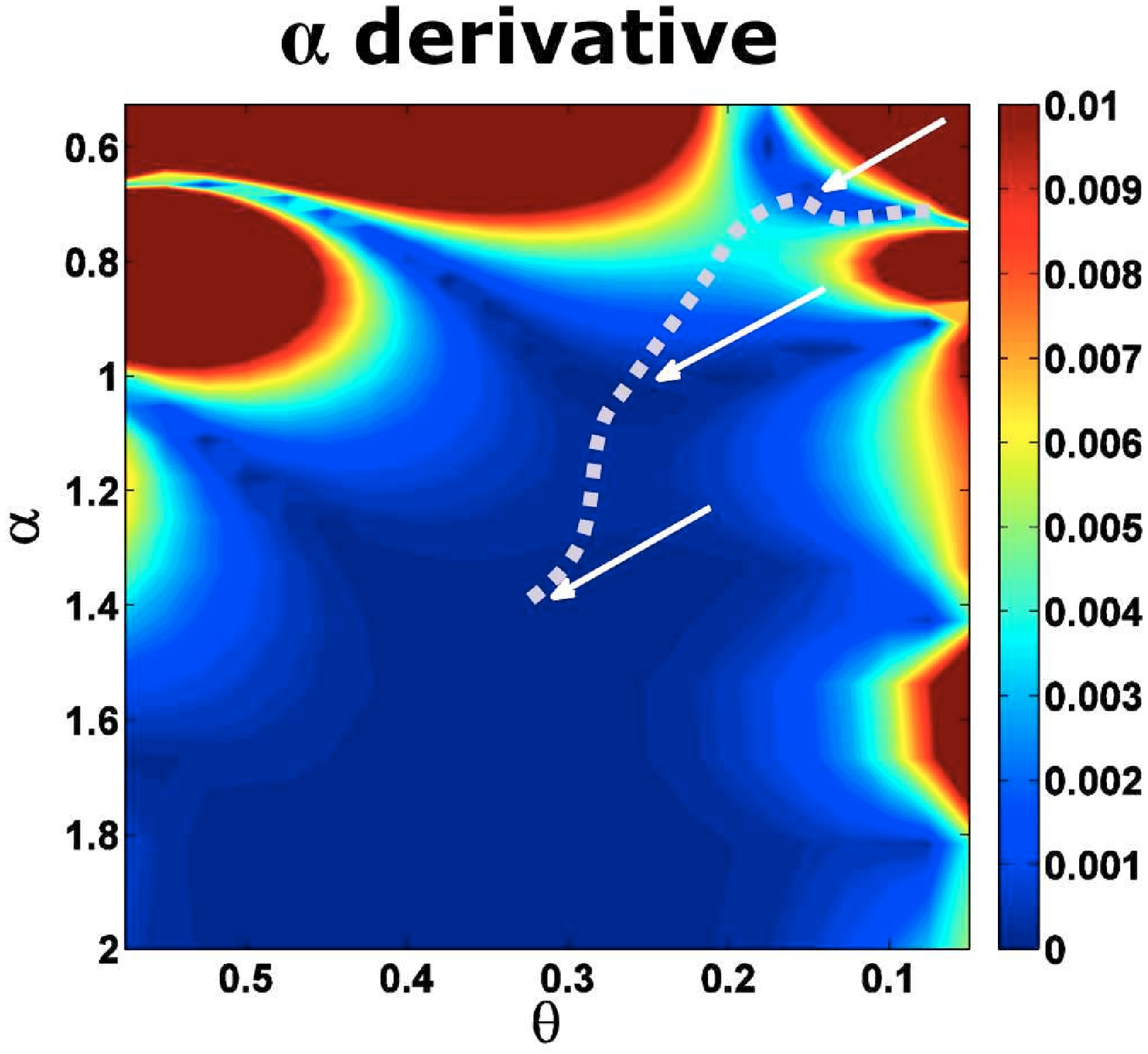}}
\caption{[Color online] 3D plot and the corresponding 2D contour plot of
$|\partial(E-i\Gamma)/\partial \theta|$ (a, c) and
$|\partial(E-i\Gamma)/\partial \alpha|$ (b, d) of the $2s^2$ helium resonance as function of $\alpha$ and $\theta$. White arrows mark the complex stationary solutions, for which the complex derivative is minimal.
These points are associated with the cusps in Fig.~\ref{cusps}. Note the narrow analytical paths that go from the real axis to the complex plane.
A gray dashed line demonstrates such a path. This path starts from a certain area in the real stabilization plot and goes to the complex plane through the three stationary points.}
\label{2D3D}
\end{figure}

In order to better understand how our analytical continuation scheme works, a thorough investigation was done using the UCS calculations. During this investigation, the complex eigenvalue associated with He ($2s^2$) resonance was calculated at 720 different scaling parameters ($\theta$ and $\alpha$).
Fig.~\ref{2D3D} represents the absolute value of this eigenvalue derivative at these points in both 3D and 2D contour plots.
Fig.~\ref{2D3D}(a) and (c) display this derivative with respect to $\theta$, whereas Fig.~\ref{2D3D}(b) and (d) display this derivative with respect to $\alpha$.
Looking at the figures, the theory behind our analytical continuation becomes clear.
First, it is obvious that for small $\theta$s, there are regions in the derivatives (with respect to $\alpha$ and $\theta$) that exhibit very large values. These values are associated with the branch points and the avoided crossings, which are forbidden areas for analytical continuations.
Yet, it also quite clear that in spite of these forbidden areas, there are analytical paths that start on the real scaling parameter axis (i.e. $\theta$= 0) and end up in stationary points in the complex plane.
Moreover, there are even paths that lead from one resonance stationary solution to another. An example of such a path is the gray dashed line in Fig.~\ref{2D3D}(c) and (d).
This path starts from region 2 in the stabilization plot of Fig.~\ref{stabilization}(a), and goes through each stationary point that was found in Fig.~\ref{cusps}.
That is, this investigation is a numerical illustration for the possibility to find analytical paths from areas in the standard stabilization calculations to the complex stationary points.

Another interesting point that can be seen in Fig.~\ref{2D3D} is that the entrance to the complex plane from the real axis always starts as a very narrow path. In fact, these passages occur only since the calculation employs a finite basis set.
For an infinite basis set the passages will close down, since as the basis set approaches completeness the number of avoided crossing increases, and the singularity area in the complex plane increases too.
That is, there is no way to analytically dilate the single eigenvalue and avoid the branch points, the passes are close.

The existence of only narrow paths in a finite basis set framework, indicates that a careful search for these analytical paths in the complex parameter plane ($\alpha$ and $\theta$) is required and still, the Pad\'{e} approximant has no problem locating these paths. However, in the absence of such analytical paths the method presented here is not applicable and one may use the analytical continuation of the characteristic polynomials as described in Ref.~\cite{mccurdy1983}.

%

\begin{figure}
\includegraphics[scale=0.33]{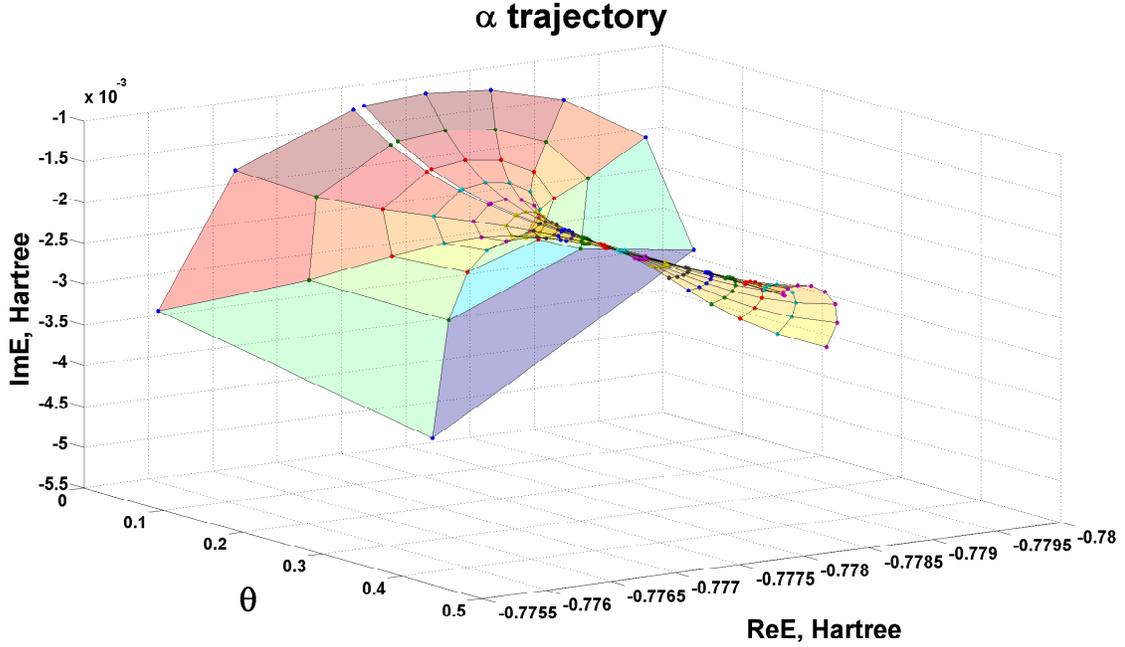}
\caption{[Color online] A 3D plot of the helium $2s^2$ complex eigenvalue as $\theta$ is varied.
This plot shows $\alpha$-trajectories at different fixed $\theta$'s. The stationary solutions (cusps in the $\alpha$ trajectories) are clearly seen for larger values of $\theta$, and it is clear that they can all be connected through certain paths.
For rather small $\theta$ a large dispersion in the complex eigenvalue is shown. However, the paths that connect the different cusps start from a relatively dense area marked by the warmer colors (red-yellow). These areas correspond to the stable part of the stabilization plot in Fig. 1(a) (the red and green region there). }
\label{swirl}
\end{figure}

Fig.~\ref{swirl} is another illustration to the information in Fig.~\ref{2D3D}.
In this figure we show a 3D plot of the helium 2$s^2$ complex eigenvalue as $\theta$ is varied, i.e., we show $\alpha$-trajectories at different fixed $\theta$s. In this figure we examine the complex eigenvalue itself, unlike in Fig.~\ref{2D3D} where the derivative of the complex eigenvalue was examined. It is clear that cusps are obtained in the $\alpha$-trajectory calculations as $\theta$ becomes larger.  However, it is also clear that there are paths connecting the different cusps and that these paths start from very specific regions. These regions are the stable parts of the $\alpha$-trajectory at small $\theta$s, where the eigenvalues are relatively close to one another. In other words, these paths go to the stable part of the stabilization plot. In Fig.~\ref{swirl} these paths are indicated in the warm colors (red-yellow), whereas in Fig.~\ref{2D3D} such a path is marked by the gray dashed line.

\clearpage

\clearpage

\subsection{Autoionization Feshbach resonance of the hydrogen atomic anion}

In order to further examine our Pad\'{e} analytical continuation scheme we calculate the 2$s^2$
Feshbach resonance of the hydrogen atomic anion.
The results are presented in Table~\ref{tbl:h_minus} where they are compared to experimental and other theoretical works. An excellent agreement with these estimations is observed, as our results are well within the experimental errors.
Furthermore, the energy position in this work is lower by 2$\times10^{-4}$ Hartree from other theoretical works, and the width is only 1.5$\times10^{-4}$ Hartree lower than Ho's \cite{ho1981H-} and Chen's \cite{chen1997H-} results,
while it is 2$\times10^{-4}$ Hartree higher than the width obtained by Bravaya et al. \cite{ksenia2013complex}.
Additional computational details and further results can be found in the supplementary information.

\begin{table}[h!]
\def\VSpace{0.25cm}
{\caption{The real part of the energy (ReE) and width ($\Gamma$) of the 2$s^2$ Feshbach resonance of H$^-$ in Hartree
\protect\label{tbl:h_minus}}}
\begin{center}
\begin{tabular}{p{3.95cm}  |p{2.5cm} p{2.cm}  }
\hline
\hline
\multicolumn{1}{c}{Ref.}&
\multicolumn{1}{|c}{-ReE }&
\multicolumn{1}{c}{$\Gamma\times10^{-3}$ }\\
\hline
\multicolumn{3}{c}{Experiment}\\
\hline
McGowan (1967)$^a$  & \quad 0.1485  & \quad 1.6 \\
                     &   \footnotesize$\pm$0.0004  &  \footnotesize$\pm$0.2  \\[5pt]
William (1976)$^b$  & \quad 0.1488     & \quad 1.65  \\
                         &   \footnotesize$\pm$0.0004  &  \footnotesize$\pm$0.2  \\[5pt]
\hline
\multicolumn{3}{c}{Theory}\\
\hline
Ho (1981)$^c$    & \quad  0.1487765  & \quad 1.731  \\
                    &   \footnotesize$\pm$0.000002  &  \footnotesize$\pm$0.0008  \\[5pt]
Chen (1997)$^d$ & \quad  0.148782  & \quad 1.72  \\
Bravaya {\it et al.} (2013)$^e$ & \quad 0.1488 & \quad 1.38 \\
{\bf Present work (Pad\'{e})} & \quad 0.14855 & \quad 1.56 \\
                            &   \footnotesize$\pm$0.00001  &  \footnotesize$\pm$0.03  \\[5pt]
\hline
\hline
\end{tabular}\\[\VSpace]
{\it a} - Ref \cite{experiment1967H-},
{\it b} - Ref \cite{experiment1976H-},
{\it c} - Ref \cite{ho1981H-},
{\it d} - Ref \cite{chen1997H-},
{\it e} - Ref \cite{ksenia2013complex}.
\end{center}
\end{table}

\clearpage


\begin{table}[h!]
\def\VSpace{0.25cm}
{\caption{The real part of the energy (ReE) and width ($\Gamma$)  of the 1$\sigma_u^2$ Feshbach resonance of H$_2$ ($R$=1.4 {\it a.u.}) in Hartree, ReE is presented with respect to the H$_2^+$ ground state energy (-0.56994 Hartree).
\protect\label{tbl:h2}}}
\begin{center}
\begin{tabular}{p{2.75cm}  |p{2.cm} p{2.cm}  }
\hline
\hline
\multicolumn{1}{c}{Method}&
\multicolumn{1}{|c}{ReE }&
\multicolumn{1}{c}{$\Gamma$}\\
\hline
Complex CI$^a$    & \quad  0.4630  & \quad 0.0272  \\
CMCSCF$^a$         & \quad  0.4638  & \quad 0.0270  \\
RF-CAP$^b$          & \quad  0.4615  & \quad 0.0227 \\
{\bf Basis1(Pad\'{e})}$^c$  & \quad  0.4619  & \quad 0.0230 \\
         &   \footnotesize$\pm$0.0002  &  \footnotesize$\pm$0.001  \\[5pt]
{\bf Basis2(Pad\'{e})}$^c$  & \quad  0.4601  & \quad 0.0231 \\
         &   \footnotesize$\pm$0.00001  &  \footnotesize$\pm$0.0004  \\[5pt]
\hline
\hline
\end{tabular}\\[\VSpace]
{\it a} - Ref \cite{mccurdy1985H2}.
{\it b} - Ref \cite{sajeev2007rf}. 
{\it c} - Present work.
\end{center}
\end{table}

\begin{figure}
\includegraphics[scale=.8]{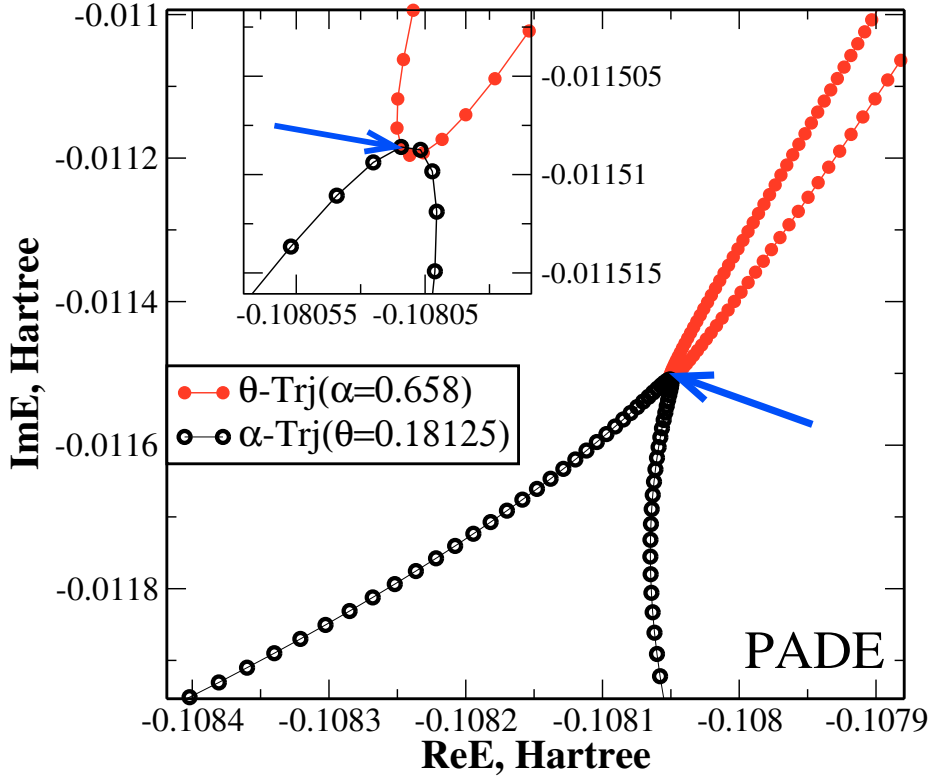}
\caption{[Color online] $\alpha$-trajectory (black) and $\theta$-trajectory (red) obtained from our analytical continuation scheme for the H$_2$ ($R=1.4$ {\it a.u.}) $1\sigma_u^2$ autoionizing resonance. In this figure an obvious cusp is seen (blue arrow), indicating a stationary point at $\theta$=0.18125 and $\alpha$=0.658. The $\alpha$-trajectory and $\theta$-trajectory overlap at the cusp, as clearly shown in the inset.}
%
\label{H2-pade}
\end{figure}

\subsection{Autoionization Feshbach resonance of the hydrogen molecule}

Calculating molecular resonance is a complicated task, which poses a challenge on the regular complex scaling methods \cite{simon1979exterior,moiseyev_corcoran1979,mccurdy1980_BO,mccurdy1980_BF,simon_morgan1981}.
Therefore, it is important to test the Pad\'{e} analytical continuation scheme presented here on a molecular system such as the hydrogen $1\sigma_u^2$ molecular resonance.
This Feshbach resonance was calculated at an internuclear distance of $R=1.4$ {\it a.u.} for which there are several calculations available for comparison.
Note that in the molecular case, when the basis functions exponents are scaled by the real factor, $\alpha$, the electronic coordinates of the Gaussian basis functions are shifted from their centers, $A_j$.
That is, every one of the Gaussian basis functions is scaled as, $G({\vec r_i-\vec A_j})\to G([{\vec r_i-\vec A_j}]/\alpha)$.

In Fig.~\ref{H2-pade} an optimal cusp for hydrogen $1\sigma_u^2$ molecular resonance is shown using Basis1 (see Section~\ref{sec:details}).
This cusp was obtained through analytical continuation of the stable region in the stabilization plot of this resonance (marked in red in Fig.~\ref{stabilization}(b)).

Table~\ref{tbl:h2} presents our results using both Basis1 and Basis2, where Basis2 is a denser basis than Basis1 (see Section~\ref{sec:details}).
We observed relatively minor effect when comparing these basis sets. The difference for the real part was in the order of 0.0018 Hartree whereas there was no difference in the imaginary part.

In Table~\ref{tbl:h2} we compare our results with other theoretical works.
Our results are in an excellent agreement with these works, particularly, with the RF-CAP \cite{sajeev2007rf}.
A good agreement with the complex CI and complex multi-configuration self consistent fields (CMCSCF) methods is also observed \cite{mccurdy1985H2}.

\section{Concluding Remarks}

In this paper, we demonstrate that a single-eigenvalue curve taken from standard stabilization plots can be analytically dilate into the complex plane. On this basis, we suggest a simple procedure that accurately locates resonance positions and widths. Hence, it opens up a new way to calculate CEPS utilizing standard electronic-structure codes. This method was successfully tested in calculating helium 2$s^2$ resonance state and was compared with explicit uniform complex scaling results yielding an excellent agreement.

In addition, an in depth analysis was presented. In this context, numerical illustration clearly indicated the presence of analytical paths from the real axis to the complex plane when using finite basis sets. These paths start from the stable region in the standard stabilization plot. They continue through a very narrow passage surrounded by huge barriers that represent the avoided crossings and branch point areas. Finally, these paths end up in valleys of stationary points.

Since these paths are analytical, they can be traced using the Pad\'{e} approximant.
In this work the Pad\'{e} approximant was generated by the Schlessinger point method \cite{Schlessinger1966}.
The input points for this analytical continuation were taken from only the stable part in the stabilization plots.
In this way, it is simple to infiltration through the analytical paths and find the resonance stationary points.
It is important to note that this procedure fails when input points are taken from the whole eigenvalue curve \cite{mccurdy1983}.

We showed that the success of the proposed approach is an outcome of our shortcomings: the mandatory use of finite basis sets in numerical calculations makes the existence of a branch point in the complex plane a rare occasion. In this case, analytical paths from the stabilization plot to the stationary point emerge.

Finally, we implemented our approach on different chemical systems. In addition to helium, the resonance energy of the hydrogen $2s^2$ atomic anion was calculated as well as the molecular Feshbach $1\sigma_u^2$ resonance of H$_2$. The results for the hydrogen $2s^2$ atomic anion were in an excellent agreement with experimental and other theoretical evaluations, whereas the results for the H$_2$ $1\sigma_u^2$ resonance were in an excellent agreement with other calculations performed explicitly in the complex plane.

In spite of the above, we do not claim that our approach avoids the need to develop and use other methods. The reason for this is the fact that a region stable enough for a successful analytical continuation is not always guaranteed.
It is recommended to first verify the conjecture of analyticity at certain geometries, particularly in the calculations of molecular autoionization resonances.
This can be done by comparing the results with other methods. As the next step, one can use the method we proposed for calculating the entire complex potential energy surface.

To sum it up, here we open a new possibility for calculating atomic and molecular autoionization resonances in a very simple manner and substantially lower computational efforts. This method has the potential to bridge between standard electronic-structure methods and calculations of CPESs, which are highly important in studying the dynamics of a molecular system for example.

\begin{acknowledgments}
This research was supported by the I-CORE Program of the Planning and Budgeting Committee, by the Israel Science Foundation grant No. 298/11, and by the Czech Ministry of
Education (grant LG13029) and institutional supports (RVO:86378271 and RVO:61388955).
Roland Lefebvre is acknowledged for giving us his code for the Pad\'{e} approximant, as presented in Ref.\cite{Schlessinger1966}.
\end{acknowledgments}

\bibliography{resonance,arik}

\end{document}